\documentclass[12pt,a4paper]{article}
\usepackage{graphicx}
\usepackage[cp1251]{inputenc}
\usepackage[english]{babel}
\usepackage{amssymb, amsfonts, amsmath}
\usepackage{amscd}
\begin{document}

\date{}

\title{Exact solutions of Maxwell equations in homogeneous spaces with group of motions $G_3(IX)$}

\author{V. V. Obukhov}

\maketitle

\quad Institute of Scietific Research and Development,
Tomsk State Pedagogical University (TSPU). Tomsk State  Pedagogical University, 60 Kievskaya St., Tomsk, 634041, Russia; \\ \quad

Laboratory for Theoretical Cosmology, International Center of Gravity and Cosmos, Tomsk State University of Control Systems and Radio Electronics (TUSUR), 36, Lenin Avenue, Tomsk, 634050, Russia

\quad


%
%
%
%

\title{Exact solutions of Maxwell equations in homogeneous spaces with group of motions $G_3(IX)$}


\section{Introduction}

 All known methods of integration of main differential equations of mathematical physics are based on complete reduction of these equations to a system of ordinary differential equations. Reduction is carried out using symmetry operators.
 For the equations of motion of  classical or quantum sample particle in external electromagnetic and gravitational fields the symmetry operators are integrals of motion.  It is known that a necessary condition for the existence of integrals of motion is the existence of spacetime symmetry given by the Killing fields.

 Thus the problem of exact integration is closely related to the study of space-time symmetry. At present two methods of exact integration of equations of motion are known. These are methods of commutative (CIM) and noncommutative (NCIM) integration.
The first method is based on the theory of complete separation of variables, and it is applicable in stackel spaces.  Stackel spaces admit  complete sets consisting of mutually commuting Killing fields.
Theory of Stackel spaces was developed in \cite{1}, \cite{2}, \cite{3}, \cite{4}, \cite{5}, \cite{6},\cite{7}. A description of the theory and a detailed bibliography can be found in \cite{8}, \cite{9} \cite{10}, \cite{11} (see also \cite{10b}). Solutions of field equations, which are still used widely in the theory of gravitation, have been constructed on the basis of Stackel spaces. These solutions are often used in the study of various effects in gravitational fields (see, for example,\cite{12}, \cite{13}, \cite{14}, \cite{15}, \cite{15a}, \cite{16}, \cite{17}, \cite{17a}, \cite{18}, \cite{19}, \cite{20}, \cite{21}, \cite{22}).

The second method (NCI method) is based on the use of noncommutative algebras of symmetry operators linear in moments and constructed using vector Killing fields. The method was proposed in \cite{23}. The development of the method and its application to gravity theory can be found in \cite{24}, \cite{25}, \cite{26}, \cite{27}.

As in stackel spaces in the spaces with a noncommutative group of motions the equations of motion of a test particle admit the complete reduction to a system of ordinary differential equations. Therefore, we will call space-time manifolds admitting  noncommutative groups $G_r, r \geq 3$ as post-Stackel spaces (PSS).

 By analogy with stackel spaces we will call the PSS non-isotropic if a group $G_r$ (or its subgroup of rank 3) acts transitive on a non-isotropic hypersurface of spacetime, or isotropic, if the hypersurface is isotropic. For non-isotropic post-stackel spaces we will also use the term "homogeneous post-stackel space (HPSS)".

The same classification problems can be considered for the PSS as for the stackel spaces.
For example, in the papers \cite{9} \cite{10}, \cite{10a} a complete classification is given for the case when the Hamilton-Jacobi equation for a charged test particle admits the complete separation of variables in the external electromagnetic field. A similar classification problem has been solved for PSS-spaces as well. In \cite{28} PSS-spaces with transitive four-parameter groups of motions are considered; in \cite{29} HPSS-spaces are considered (see also \cite{29a}); in \cite{30} PSS-spaces with groups acting on isotropic hypersurfaces of transitivity are considered. PSS-spaces with four-parameter groups of motions are considered in \cite{31}, provided that these groups have transitive three-parameter subgroups.
Thus, one has found the potentials of all admissible electromagnetic fields, for which the Hamilton-Jacobi and Klein-Gordon-Fock  equations have algebras of symmetry operators given by groups of motions of post-stackel spaces. It was proved, that the Klein-Gordon-Fock equation admits the algebra of symmetry operators given by groups of motions of PSS if and only if the Hamilton-Jacobi equations admits the appropriate algebra of integrals of motion.

 Next classification problem is the classification of electrovacuum solutions of the Einstein-Maxwell equations for the case, when CIM and NCIM methods are applicable. During the century-long history of General relativity, many exact solutions of the vacuum and electrovacuum Einstein equations have been found (see, for example,\cite{31a} ).
 Nevertheless, this problem has not lost its relevance up to now. The main purpose of the classification is not so much to find new exact solutions, as to list all gravitational and electromagnetic fields, in which equations of motion of test particles can be exactly integrated or at least reduced to systems of ordinary differential equations. This problem divided into two stages.

 At the first stage all non-equivalent classes of solutions of the vacuum Maxwell equations for the potentials of admissible electromagnetic fields are found. At the second stage the obtained classification is used to classify the corresponding electrovacuum spaces. Historically, for Stackel spaces this problem was solved before the problem of the first stage (see the bibliography given in \cite{9}, \cite{10}, \cite{10a}).
 The present article is devoted to solving the first stage of this classification problem. All non-equivalent solutions of empty Maxwell equations in homogeneous spaces of type IX according to Bianchi's classification are found.

\section{Admissible  electromagnetic fields in homogeneous spaces}

 There are two definitions of homogeneous spaces. According to the first  a spacetime \quad $V_4$ \quad is homogeneous if its subspace \quad $V_3$, \quad endowed with the Euclidean space signature, admits coordinate transformations (forming the group $G_3(N)$ of motions of spaces $V_4$), that allow to connect any two points in \quad $V_3$ \quad (see \cite{32}). This definition directly implies that metric of the  $V_4$ in the semi-geodesic coordinate system \quad $[u^i]$ \quad  can be represented as follows:
\begin{equation}\label{1}
ds^2 = -{du^0}^2 + \eta_{ab}l_{\alpha}^a l_{\beta}^b du^\alpha du^\beta,  \quad g_{ij}=-\delta_i^0\delta_j^0 + \delta_i^a\delta_j^b e^a_\alpha e^b_\beta \eta_{ab}(u^0), \quad det|\eta_{ab}|>0 \quad e^a_{\alpha, 0}=0.
\end{equation}
The coordinate indices of the variables of the semi-geodesic coordinate system are denoted by lower case Latin letters:\quad $i, j, k = 0, 1 \dots 3$.\quad The coordinate indices of the variables of the local coordinate system on the hypersurface \quad $ V_3$ \quad are denoted by lower case Greek letters:\quad $\alpha, \beta, \gamma=1, \dots 3.$\quad the time variable is denoted by a 0 index. Group indices and indices of nonholonomic frame are denoted by \quad $a, b, c = 1, \dots 3 $.\quad Summation is performed over repeated upper and lower indices within the index range.

The 1-form \quad $e^a_\alpha du^\alpha \quad $
is invariant under the acting of the group $G_3(N)$. The vectors of the frame $e^a_\alpha$ define a non-holonomic coordinate system in $V_3$. The dual triplet of vectors
$$
e_a ^\alpha, \quad e_a ^\alpha e^b_\alpha =\delta_a^b, \quad  e_a ^\alpha e^a_\beta = \delta^\alpha_\beta
$$
defines the operators of the group algebra:
\begin{equation}\label{2a}
\hat{Y}_a= e_a ^\alpha \partial_a, \quad [\hat{Y}_a,\hat{Y}_b] = C_{ab}^c \hat{Y}_c.
\end{equation}
According to another definition, space-time \quad $V_4$ \quad is homogeneous if it admits a three-parameter group of motions \quad $G_3(N)$, \quad whose hypersurface \quad $V_3$ \quad  of transitivity has the Euclidean space signature.
The Killing vector fields $\xi_a^\alpha $ \quad and their dual vector fields \quad $\xi^a_\alpha$ \quad form another frame of the space $V_3$ and another representation of the algebra of the group $G_3$.
The vector fields \quad $\xi_a^\alpha$ \quad satisfy the Killing equations:
\begin{equation}\label{2}
g^{\alpha\beta}_{,\gamma} \xi_a ^\gamma = g^{\alpha \gamma}\xi^\beta_{a,\gamma}+g^{\beta\gamma}\xi^\alpha_{a,\gamma},
\end{equation}
and sets the infinitesimal group operators of the algebra $G_3$
\begin{equation}\label{3}
\hat{X}_a= \xi_a ^\alpha \partial_\alpha, \quad [\hat{X}_a,\hat{X}_b] = C_{ab}^c \hat{X}_c.
\end{equation}
Let us consider electromagnetic field with potential $A_i$.
For a charged test particle, moving in this external electromagnetic field, it has been proved, that the Hamilton-Jacobi equation:
\begin{equation}\label{4}
g^{ij}P_iP_j=m, \quad P_i=p_i+A_i.
\end{equation}
and the Klein-Gordon-Fock equation:

\begin{equation}\label{1f}
\hat{H}\varphi = (g^{ij}\hat{P}_i\hat{P}_j)\varphi=m^2\varphi, \quad \hat{P}_k=\hat{p}_k+A_k.
\end{equation}
admit the integrals of motion, which are given by Killing vectors:
$$
\tilde{X}_\alpha=\xi_\alpha^i p_i \quad (or \quad \hat{\tilde{X}}_\alpha=\xi_\alpha^i \hat{p}_i),
$$
if and only if the conditions:
\begin{equation}\label{8}
\xi^\alpha_a(\tilde{\mathbf{A}})_{,\alpha}= C^c_{ab}\tilde{\mathbf{A}} \end{equation}
are satisfied (see papers \cite{29}).
Here \quad $ p_i=\partial_i\varphi $; \quad $\hat{p}_k=-\imath\hat{\nabla}_k;\quad  $  ($ \hat{\nabla}_k$ is the covariant derivative operator corresponding to the partial derivative operator -\quad   $\hat{\partial}_i$   \quad   in the coordinate field \quad   $u^i), \quad \varphi $\quad is a scalar function of particle with mass $\quad m $; $\quad \tilde{\mathbf{A}}_a = \xi^\alpha_a A_\alpha.$

The electromagnetic field whose potential satisfies condition \eqref{8} is called admissible.
All admissible electromagnetic fields for groups of motion \quad $G_r(N)\quad ( r\leq 4)$, \quad acting transitively on hypersurfaces of the spacetime, have been found in \cite{29}, \cite{30}, \cite{31}.

Solutions of the set of equations \eqref{8} for HPSS of type $IX$ have the form:
\begin{equation}\label{8a}
A_{\alpha}=\alpha_a(u^0) l_{\alpha}^a  \Rightarrow \mathbf{A}_a = l^{\alpha}_a A_\alpha = \alpha_a(u^0).
\end{equation}
To prove this let's find the  frame vector. We will use the metric tensor of $IX$-type space by Bianchi, found in Petrov's book \cite{33}. As it is known, the Bianchi type IX metric contains as a special case the space of constant positive curvature and therefore is of special interest for cosmology.
\begin{equation}\label{9f}
{ds}^2 = {du^1}^2[a_{11}-(a_{12}\cos{2u^3}+a_{22}\sin{2u^3})]+2{du^1du^3}((a_{13}\cos{u^3}-
a_{23}\sin{u^3})+
\end{equation}$$
+2{du^1du^2}[(a_{13}\cos{u^3}-a_{23}\sin{u^3})\cos{u^1}+(a_{12}\cos{2u^3}-a_{22}\sin{2u^3})\sin{u^1}]
$$
$$
+{du^2}^2[a_{33}{\cos{u^1}}^2+(a_{23}\cos{u^3}+a_{13}\sin{u^3})\sin{2u^1}+
(a_{12}\sin{2u^3}+a_{22}\cos{2u^3}+a_{11}){\sin{u^1}}^2  ]
$$
$$
2{du^2du^3}(a_{33}\cos{u_1}+(a_{23}\cos{u^3}+a_{13}\sin{u^3})\sin{u^1})+{du^3}^2a_{33}+
e{du^0}^2.$$
$a_{ab}$  are arbitrary functions on $u^0$.

 To obtain the functions \quad $l^\alpha_a$ \quad, it is sufficient to consider the components \quad$g_{13},g_{23}\quad $ \quad from the system \eqref{2}. The solution has the form:
$$
l_\alpha^a = \delta_\alpha^p l_p^a(u^1,u^3)+\delta_\alpha^3 \delta^a_3\quad.
$$
From the equations:
$$
g_{13} = a_{13}\cos{u^3}-a_{23}\sin{u^3} =\eta_{3a}l^a_1, \quad g_{23} = a_{33}\cos{u_1}+(a_{23}\cos{u^3}+a_{13}\sin{u^3})\sin{u^1}= \eta_{3a}l^a_1
$$
it follows:
\begin{equation}\label{1c}
 l^a_\alpha=\begin{pmatrix} \cos{u^3} &-\sin{u^3}  & 0 \\ \sin{u^1}\sin{u^3} & \sin{u^1}\cos{u^3} & \cos{u^1} \\0 & 0 & 1\end{pmatrix},
  l_a^\alpha=\begin{pmatrix} \cos{u^3} &\frac{\sin{u^3}}{\sin{u^1}}  & -\frac{\cos{u^1}\sin{u^3}}{\sin{u^1}} \\-\sin{u^3} &
 \frac{\cos{u^3}}{\sin{u^1}} & -\frac{\cos{u^1}\cos{u^3}}{\sin{u^1}} \\ 0 & 0 & 1  \end{pmatrix},
\end{equation}
$l^a_\alpha l^\alpha_b = \delta^a_b. \quad $

The lower index numbers the lines. One can show  that the vector fields \eqref{1c} satisfy the equations \eqref{1}, \eqref{2a}: We present the components of the vectors $\xi_a^\alpha$ in the form of a matrix:
$$
  ||\xi_a^\alpha||=\begin{pmatrix} 0 & 1 & 0 \\ \cos{u^2} &-\frac{\cos{u^1}\sin{u^2}}{\sin{u^1}} & \frac{\sin{u^2}}{\sin{u^1}} \\ -\sin{u^2} &
 -\frac{\cos{u^1}\cos{u^2}}{\sin{u^1}} & \frac{\cos{u^2}}{\sin{u^1}} \end{pmatrix}$$
The components $\tilde{\mathbf{A}}_{\alpha}$ can be expressed through $\mathbf{A}_{\alpha}$ as follows:
$$
 \tilde{\mathbf{A}}_{a}=Z_a^b\mathbf{A}_b,
$$
where
$$
    ||Z_a^b=\xi^{\alpha}_a l^b_{\alpha}||=\begin{pmatrix} \sin{u^1}\sin{u^3}  & \sin{u^1}\cos{u^3} & \cos{u^1} \\(\cos{u^2}\cos{u^3}-\sin{u^2}\sin{u^3}\cos{u^1})  & -(\cos{u^2}\sin{u^3}+\sin{u^2}\cos{u^3}\cos{u^1}) & \sin{u^1}\sin{u^2}\\-(\sin{u^2}\cos{u^3}+\cos{u^2}\sin{u^3}\cos{u^1})  & (\sin{u^2}\sin{u^3}-\cos{u^2}\cos{u^3}\cos{u^1}) & \cos{u^2}\sin{u^1} \end{pmatrix}.
 $$
It can be shown by direct calculation that the elements of the matrix $Z_a^b$ satisfy the equation:
\begin{equation}\label{9}
Z^b_{a|c}= C_{ca}^{a_1} Z_{a_1}^b,\quad |a = l^{\alpha}_a \partial_\alpha.
\end{equation}
Therefore, the equation \eqref{8} can be reduced to the form:
\begin{equation}\label{2f}
\xi_a^\alpha\mathbf{A}_{b,\alpha} = 0 \Rightarrow \mathbf{A}_{a}  = \alpha_a(u^0).
\end{equation}

\section {Maxwell's equations with zero electromagnetic field sources in a homogeneous spacetime}

All exact solutions of vacuum Maxwell equations for solvable groups have been found in the papers \cite{34}, \cite{35}. In the present paper the problem is solved for the group $G_3(IX)$.

We will use the first definition of homogeneous spaces. Note, that for the space-time with the groups of motions $G_3(I)-G_3(VI),  G_3(IX)$  both definitions are equivalent

Consider the Maxwell equations with zero electromagnetic field sources in homogeneous space in the presence of an electromagnetic field invariant with respect to the group $G_r$:
\begin{equation}\label{13}
\frac{1}{\sqrt{-g}}(\sqrt{-g}F^{ij})_{,j} = 0,
\end{equation}
The metric tensor is defined by relations \eqref{1}, the electromagnetic potential by the relations \eqref{8}.
When $i=0$, from the set of equations \eqref{13} it follows:
\begin{equation}\label{14}
\quad\frac{1}{\sqrt{-g}}(\sqrt{-g}g^{\alpha\beta}F_{0\beta})_\alpha = \frac{1}{l}(l l^{\alpha}_{a}\eta^{ab}\dot{\alpha}_{b})_{,\alpha} = \eta^{ab}\rho_a\dot{\alpha}_b =0.
\end{equation}
Here it is denoted
$g=-\det||g_{\alpha\beta}||=-(\eta l)^2,$ \quad where \quad $\eta^2 = \det||\eta_{\alpha\beta}||, \quad l=\det||l_{\alpha}^a||,\quad \rho_a  = l^\alpha_{a,\alpha} + l_{|a}/l,$ \quad the dots means the time derivatives.
Let \quad $i=\alpha$. \quad Then from the equation \eqref{13} it follows:
\begin{equation}\label{16}
\frac{1}{\eta}(\eta g^{\alpha\beta}F_{0\beta})_{,0} = \frac{1}{l}(lg^{\nu\beta}g^{\alpha\gamma}F_{\beta \gamma})_{,\nu} \Rightarrow \frac{1}{\eta}(\eta \eta^{ab}l^\alpha_a \dot{\alpha}_{b})_{,0}=\frac{1}{l}(l l_a^\nu l^\beta_b \eta^{ab} l_{\tilde{a}}^{\alpha} l_{\tilde{b}}^{\gamma} \eta^{\tilde{a}\tilde{b}}F_{\beta\gamma})_{,\nu}\Rightarrow
\end{equation}
\begin{equation}\label{16a}
(\eta\eta^{ab}\dot{\alpha}_b)_{,0} = \frac{\eta l_\alpha^a}{l}(l l^\beta_b l_{\tilde{a}_1}^{\alpha} l_{\tilde{b}}^{\gamma}F_{\beta\gamma})_{|{a_1}}\eta^{{a_1}b}\eta^{\tilde{a}\tilde{b}}.
\end{equation}
Let us find components of $F_{\alpha\beta}$, using  the relations \eqref{8a}.
\begin{equation}\label{17}
F_{\alpha\beta} = (l^a_{\beta,\alpha}-l^a_{\beta,\alpha})\alpha_a = l^c_\beta l^\gamma_c l^d_\alpha l^\nu_d(l^a_{\gamma,\nu}-l^a_{\nu,\gamma})\alpha_a=l^b_\beta l^a_\alpha l^c_\gamma  (l^\gamma_{a|b}-l^\gamma_{b|a})\alpha_c = l^b_\beta l^a_\alpha C^c_{ba}\alpha_c.
\end{equation}
Then
\begin{equation}\label{18}
(lF^{\alpha\beta})_{,\beta} = \eta^{ab}\eta^{\tilde{a}\tilde{b}}C^d_{\tilde{b}b}\alpha_d ((ll^\alpha_a)_{|\tilde{a}} + ll^\alpha_al^\gamma_{\tilde{a},\gamma}).
\end{equation}
Structural constants of a  group \quad $G_3$ \quad can be present in the form:
\begin{equation}\label{19}
C^c_{ab} = C^c_{12}\varepsilon^{12}_{\tilde{a}\tilde{b}} + C^c_{13}\varepsilon^{13}_{\tilde{a}\tilde{b}} + C^c_{23}\varepsilon^{23}_{\tilde{a}\tilde{b}},\quad
\varepsilon^{AB}_{ab} = \delta^A_a\delta^B_b - \delta^A_b\delta^B_a.
\end{equation}
Using the notations:
$$
\sigma_1 = C^a_{23}\alpha_a, \quad \sigma_2 = C^a_{31}\alpha_a, \quad \sigma_3 = C^a_{12}\alpha_a, \quad
$$
$$
\gamma_1=\sigma_1\eta_{11}+\sigma_2\eta_{12}+\sigma_3\eta_{13}, \quad \gamma_2=\sigma_1\eta_{12}+\sigma_2\eta_{22}+\sigma_3\eta_{23}, \quad \gamma_3=\sigma_1\eta_{13}+\sigma_2\eta_{23}+\sigma_3\eta_{33},
$$
 let us reduce Maxwell's equations \eqref{13} to the form:
\begin{equation}\label{19}
\eta(\eta^{ab}\dot{\alpha}_b)_{,0} = \delta^a_1(\gamma_1(C^1_{32}) - \gamma_2(C^1_{31} +\rho_3) + \gamma_3(C^1_{21} +\rho_2)) + \delta^a_2(\gamma_1(C^2_{32} +\rho_3) +
\end{equation}
$$
\gamma_2C^2_{13} - \gamma_3(C^2_{12} \rho_1)) +\delta^a_3(-\gamma_1(C^3_{23} +\rho_2) + \gamma_2(C^3_{13} +\rho_1) + \gamma_3C^3_{21}),
$$
The order of the equations \eqref {19} can be decreased by introducing a new independent functions:
\begin{equation}\label{20}
 \beta_a = \beta^a = \eta\eta^{ab}\dot{\alpha}_b \quad \Rightarrow \quad\eta \dot{\alpha}_a = \eta_{ab}\beta^b.
\end{equation}
Let us consider the  Maxwell equations for the group $G_3(IX)$. As in this case \quad \quad non zero  structural constants are following:
$$C^3_{12} = C^2_{31} = C^1_{23} = 1, $$
functions $\quad \sigma_a, \gamma_1$  \quad have the form:
$$
 \sigma_1=\alpha_1, \quad \sigma_2 = \alpha_2, \quad \sigma_3 = \alpha_3. \quad
$$
$$
\gamma_1=\alpha_1\eta_{11}+\alpha_2\eta_{12}+\alpha_3\eta_{13}, \quad \gamma_2=\alpha_1\eta_{12}+\alpha_2\eta_{22}+\alpha_3\eta_{23}, \quad \gamma_1=\alpha_1\eta_{13}+\alpha_2\eta_{23}+\alpha_3\eta_{33}.
$$
Using these relations, we obtain Maxwell's equations \eqref{14}, \eqref{19} as a system of linear algebraic equations on the unknown functions \quad $n_{ab}$:
\begin{equation}\label{7f}
  n_{ab} =\frac{\eta_{ab}}{\eta} \Rightarrow \eta = \frac{1}{\det{n_{ab}}}.
\end{equation}

\begin{equation}\label{2c}
\hat{W}\hat{n} = \hat{\omega},
\end{equation}
where
\begin{equation}\label{3c}
\hat{W}=\begin{pmatrix} \alpha_1 & \alpha_2 & \alpha_3 & 0 & 0 & 0
\\ \beta_1 &\beta_2 &\beta_3 & 0 & 0 & 0 \\ 0 & \alpha_1 & 0 &\alpha_2 &\alpha_3 & 0 \\ 0 & \beta_1 & 0 &\beta_2 &\beta_3 & 0 \\ 0 & 0 & \alpha_1 & 0 & \alpha_2 &\alpha_3
\\ 0 & 0 & \beta_1 & 0 &\beta_2 & \beta_3 \end{pmatrix},
\end{equation}
$$
\hat{n}^T = ( n_{11}, n_{12}, n_{13}, n_{22}, n_{23}, n_{33}) ; \quad
\hat{\omega}^T = (-\dot{\beta}_{1}, \dot{\alpha}_{1}, -\dot{\beta}_{2}, \dot{\alpha}_{2}, -\dot{\beta}_{3}, \dot{\alpha}_{3}),
$$
index T means the transposition of a matrix.
Let us find the algebraic complement of the matrix $\hat{W}:$
\begin{equation}\label{4c}
\hat{V}=\begin{pmatrix} \beta_1 V^2_1 & -\alpha_1V^2_1 &\beta_2 V^2_1 &-\alpha_2V^2_1 & \beta_3 V^2_1 & -\alpha_3V^2_1
\\\beta_1 V_1V_2 & -\alpha_1V_1V_2 &\beta_2 V_1V_2 &-\alpha_2V_1V_2 & \beta_3 V_1V_2 & -\alpha_3V_1V_2\\\beta_1 V_1V_3 & -\alpha_1V_1V_3 &\beta_2 V_1V_3 &-\alpha_2V_1V_3 & \beta_3 V_1V_3 & -\alpha_3V_1V_3 \\ \beta_1 V^2_2 & -\alpha_1V^2_2 &\beta_2 V^2_2 &-\alpha_2V^2_2 & \beta_3 V^2_2 & -\alpha_3V^2_2 \\ \beta_1 V_2V_3 & -\alpha_1V_2V_3 &\beta_2 V_2V_3 &-\alpha_2V_2V_3 & \beta_3 V_2V_3 & -\alpha_3V_2V_3
\\\beta_1 V^2_3 & -\alpha_1V^2_3 &\beta_2 V^2_3 &-\alpha_2V^2_3 & \beta_3 V^2_3 & -\alpha_3V^2_3\end{pmatrix}
\end{equation}
As $\hat{W}$ is singular matrix, $\hat{V}$ is the annulling matrix for $\hat{W}$:
\begin{equation}\label{8f}
\hat{V}\hat{W}=0.
\end{equation}
Therefore, one of the equations from the system \eqref{2c} can be replaced by the equation:
\begin{equation}\label{5c}
\delta^{ab}(\dot{\alpha}_a \dot{\alpha}_b +\dot{\beta}_a\dot{\beta}_b) \Rightarrow \delta^{ab}(\alpha_a\alpha_b + \beta_a\beta_b) =c^2 =const.
\end{equation}
Depending on the rank of the matrix $\hat{W}$, one or more functions $n_{ab}$  are independent.
The remaining functions $n_{ab}$ can be expressed through them and through the functions $\alpha_a, \beta_a$.
For classification it is necessary to find non-equivalent solutions of the system \eqref{2c}. Obviously, this system are symmetric with respect to the transposition \quad $l_1^\alpha \leftrightarrow l_2^\alpha $. Therefore the reference indices $a=1$ and $a=2$ can be interchanged. Taking this observation into account, let us consider all non-equivalent options.

\section{Solutions of Maxwell equations}

1.  \quad $a_1V_1 \ne 0 \Rightarrow$ the minor $\hat{W}_{12}$ and its inverse matrix $\hat{\Omega}=\hat{W}_{12}^{-1}$ have the form:
\begin{equation}\label{6c}
\hat{W}_{12}=\begin{pmatrix} \alpha_2 & \alpha_3 & 0 & 0 & 0
 \\ \alpha_1 & 0 &\alpha_2 &\alpha_3 & 0 \\ \beta_1 & 0 &\beta_2 &\beta_3 & 0 \\ 0 & \alpha_1 & 0 & \alpha_2 & \alpha_3
\\0 & \beta_1 & 0 &\beta_2 & \beta_3 \end{pmatrix},
\end{equation}
\begin{equation}\label{7.c}
\hat{\Omega}_1=\begin{pmatrix}-\frac{V_2}{\alpha_1V_1} & -\frac{\alpha_3\beta_2 }{\alpha_1V_1} & \frac{\alpha_2 \alpha_3 }{\alpha_1V_1} & -\frac{\alpha_3\beta_3}{\alpha_1V_1} & \frac{\alpha^2_3}{\alpha_1V_1}
 \\
-\frac{V_3}{\alpha_1V_1} & \frac{\alpha_2\beta_2}{\alpha_1V_1} & -\frac{\alpha_2^2 }{\alpha_1V_1} &\frac{\alpha_2 \beta_3}{\alpha_1V_1} & -\frac{\alpha_2\alpha_3}{\alpha_1V_1}
\\
-\frac{V_2^2}{\alpha_1V_1^2} &\frac{(\alpha_3 \beta_1V_1 -\alpha_2 \beta_3V_3)}{\alpha_1V_1^2} & \frac{\alpha_3(\alpha_2 V_2-\alpha_1 V_1)}{\alpha_1V_1^2} &-\frac{\alpha_3\beta_3 V_2}{\alpha_1V_1^2} & \frac{\alpha_2^2V_2}{\alpha_1V_1^2}
\\
 -\frac{V_2V_3 }{\alpha_1V_1^2} & \frac{\alpha_2\beta_2V_2}{\alpha_1V_1^2} &-\frac{\alpha^2_2 V_2}{\alpha_1V_1^2} &-\frac{\alpha_3\beta_3`V_3}{\alpha_1V_1^2} & \frac{\alpha_3^2 V_3}{\alpha_1V_1^2}
\\
-\frac{V_3^2}{\alpha_1V_1^2} & \frac{\alpha_2\beta_2V_3}{\alpha_1V_1^2}&-\frac{\alpha^2_2 V_3 }{\alpha_1V_1^2}&\frac{(\alpha_3 \beta_2V_3-\alpha_2 \beta_1V_1)}{\alpha_1V_1^2} & \frac{\alpha_2 (\alpha_1V_1 - \alpha_3 V_3)}{\alpha_1V_1^2}\end{pmatrix}
\end{equation}
Then the solution of equation \eqref{2c} can be represented as:
\begin{equation}\label{8c}
\hat{n}_1 = \hat{\Omega}_1\hat{\omega}_1,
\end{equation}
were
$$
\hat{n}_1^{T} = (n_{12}, n_{13}, n_{22}, n_{23}, n_{33}) ; \quad
\hat{\omega}_1^{T} =  (-(\dot{\beta}_{1}+\alpha_1 n_{11}), -\dot{\beta}_{2}, \dot{\alpha}_{2}, -\beta_{3}, \dot{\alpha}_{3}),
$$
Function \quad $n_{11},$ \quad as well as the functions \quad $\alpha_a, \quad \beta_a$ \quad are arbitrary functions of \quad $u^0$, \quad that obey the condition \eqref{5c}.

\quad

2. \quad $\alpha_2V_1 \ne 0, \Rightarrow  \alpha_1=0 \Rightarrow$ the minor $\hat{W}^{-1}_{14}$ and its inverse matrix $\hat{\Omega}_2 = \hat{W}^{-1}_{14}$ have the form:
\begin{equation}\label{9c}
\hat{W}_{14}=\begin{pmatrix} \alpha_2 & \alpha_3 & 0 & 0 & 0
 \\ \beta_2 & \beta_3 & 0 & 0 & 0 \\ 0 & 0 &\alpha_2 &\alpha_3 & 0 \\ 0 & 0 & 0 &\alpha_2 & \alpha_3
 \\0 & \beta_1 & 0 &\beta_2 & \beta_3 \end{pmatrix},  \quad
\hat{\Omega}_2=\begin{pmatrix}
  \frac{\beta_3}{ V_1} & -\frac{\alpha_3}{V_1} & 0 & 0 & 0 \\
 -\frac{\beta_2}{V_1}& \frac{\alpha_2}{V_1} & 0 &0& 0 \\

\frac{a_3^2\beta_1\beta_2}{\alpha_2 V_1^2} &-\frac{\alpha_3^2\beta_1}{ V_1^2}&  \frac{1}{\alpha_2 } &-\frac{\alpha_3\beta_3}{\alpha_2 V_1} & \frac{a_3^2}{\alpha_2 V_1} \\

 -\frac{a_3\beta_1\beta_2}{ V_1^2} &\frac{\alpha_2\alpha_3\beta_1}{V_1^2}& 0 &\frac{\beta_3}{V_1} & -\frac{a_3}{ V_1}\\

\frac{a_2\beta_1\beta_2}{V_1}&-\frac{\alpha_2^2\beta_1}{V_1} &  0 &-\frac{\beta_2}{V_1} & \frac{\alpha_2}{V_1}\end{pmatrix}
\end{equation}
Solution of the equation \eqref{2c} can be represented as:
\begin{equation}\label{11c}
\hat{n}_2 = \hat{\Omega}\hat{\omega}_2,
\end{equation}
were
$$
\hat{n}_2 ^{T} = (n_{12}, n_{13}, n_{22}, n_{23}, n_{33}) ; \quad
$$
$$\hat{\omega}_2 = (-\dot{\beta}_{1}, -\beta_1n_{11}, -\dot{\beta}_{2},  -\dot{\beta}_{3}, \dot{\alpha}_{3})
$$
Function \quad $n_{11},$ \quad as well as the functions \quad $\alpha_a, \quad \beta_a$ \quad are arbitrary functions of \quad $u^0$, \quad that obey the condition \eqref{5c}.

\quad

3. \quad $a_3V_1 \ne 0, \Rightarrow  a_1=a_2=0 \Rightarrow$ the minor $\hat{W}^{-1}_{16}$ and its inverse matrix $\hat{\Omega}_3 = \hat{W}^{-1}_{16}$ have the form:
\begin{equation}\label{9c}
\hat{W}_{16}=\begin{pmatrix} 0 & a_3 & 0 & 0 & 0
 \\ \beta_2 & \beta_3 & 0 & 0 & 0 \\ 0 & 0 & 0 & a_3 & 0 \\ \beta_1 & 0 & \beta_2 &\beta_3 & 0
\\0 & 0 & 0 &0 & a_3 \end{pmatrix},  \quad \hat{\Omega}_3=\begin{pmatrix}-\frac{\beta_3}{a_3\beta_2} & \frac{1}{\beta_3} & 0 & 0 & 0 \\
\frac{1}{a_3} & 0 & 0 &0 & 0\\
\frac{\beta_1\beta_3}{a_3\beta_2^2} &-\frac{\beta_1}{\beta_2^2} & -\frac{\beta_3}{\beta_2 a_3} &\frac{1}{\beta_2} & 0 \\
 0 &0 &\frac{1}{a_3} & 0 & 0
\\
0 & 0 &0 & 0 & \frac{1}{a_3}\end{pmatrix}
\end{equation}
Then the solution of equation \eqref{2c} can be represented as:
\begin{equation}\label{8c}
\hat{n}_3 = \hat{\Omega}_3\hat{\omega}_3,
\end{equation}
were
$$
\hat{n}_3^{T} = ( n_{12}, n_{13}, n_{22}, n_{23},n_{33}) ; \quad
$$
$$
\hat{\omega}_3^{T} = (-\dot{\beta}_{1}, -\beta_1 n_{11}, -\dot{\beta}_{2}, 0, -\dot{\beta}_{3})
$$
Function \quad $n_{11},$ \quad as well as the functions \quad $\alpha_3, \quad \beta_a$ \quad are arbitrary functions of \quad $u^0$, \quad that obey the condition \eqref{5c}.

\quad

4. \quad $a_1V_3 \ne 0$. $\Rightarrow V_1=V_2=0,$ \quad otherwise, we get a solution equivalent to the previous ones. As \quad $V_3\ne 0 \Rightarrow$ \quad $\alpha_3=\beta_3=0.$\quad The minor $\hat{W}_{62}$ and its inverse matrix $\hat{\Omega}_4 = \hat{W}^{-1}_{62}$ have the form:
\begin{equation}\label{9c}
\hat{W}_{26}=\begin{pmatrix} \alpha_1 & \alpha_2 & 0 & 0 & 0
 \\ 0 & \alpha_1 & 0 & a_2 & 0 \\ 0 & \beta_1 & 0 & \beta_2 & 0 \\ 0 & 0 & \alpha_1 & 0 & \alpha_2
\\0 & 0 & \beta_1 & 0 & \beta_2 \end{pmatrix},  \quad \hat{\Omega}_4=
\begin{pmatrix} \frac{1}{\alpha_1} & -\frac{\alpha_2\beta_2}{\alpha_1V_3} & \frac{\alpha_2^2}{\alpha_1V_3} & 0 & 0
 \\ 0 & \frac{\beta_2}{V_3} & -\frac{\alpha_2}{V_3} & 0 & 0 \\ 0 & 0 & 0 & \frac{\beta_2}{V_3} & -\frac{\alpha_2}{V_3} \\ 0 & -\frac{\beta_1}{V_3} & \frac{\alpha_1}{V_3} & 0 & 0
\\0 & 0 & 0 & -\frac{\beta_1}{V_3} & \frac{\alpha_1}{V_3} \end{pmatrix}
\end{equation}
Then the solution of equation \eqref{2c} can be represented as:
\begin{equation}\label{8c}
\hat{n}_4 = \hat{\Omega}_4\hat{\omega}_4.
\end{equation}
were
$$
\hat{n}_4^{T} = (n_{11}, n_{12}, n_{13}, n_{22}, n_{23}) ; \quad
$$
$$
\hat{\omega}_4^{T} = (-\dot{\beta}_{1}, -\dot{\beta}_{2}, \dot{\alpha}_{2}, 0, 0).
$$
Function \quad $n_{33},$ \quad as well as the functions \quad $\alpha_1, \quad \alpha_2 \quad \beta_a$ \quad are arbitrary functions of \quad $u^0$, \quad that obey the condition \eqref{5c}.

\quad

5.\quad $V_a =0 $.\quad Let us represent the system of Maxwell equations in the form:
$$\hat{Q}_I\hat{n}_I = \hat{\omega}_I$$
were
\begin{equation}\label{9c}
\hat{Q} = \begin{pmatrix}\alpha_1 & \alpha_2 & \alpha_3 & 0 & 0 & 0 \\ 0 &  \alpha_1 & 0 & \alpha_2 & \alpha_3 &0
\\0 & 0 & \alpha_1 & 0 & \alpha_2 & \alpha_3
 \\ \beta_1 & \beta_2 & \beta_3 & 0 & 0 & 0 \\0 & \beta_1 & 0 & \beta_2 & \beta_3 & 0
\\0 & 0 & \beta_1 & 0 & \beta_2 & \beta_3\end{pmatrix},  \quad
\end{equation}
$$
\hat{\omega}_I = (\hat{\omega}_\beta, \hat{\omega}_\alpha); \quad \hat{\omega}_\beta =-(\dot{\beta}_1,\dot{\beta}_2,\dot{\beta}_3), \quad \hat{\omega}_\alpha = (\dot{\alpha}_1,\dot{\alpha}_2,\dot{\alpha}_3)
$$
$$
\hat{n}_I = (\hat{n}_\alpha, \hat{n}_\beta); \quad \hat{n}_\alpha =(n_{11},n_{12},n_{13}), \quad \hat{n}_\beta =(n_{22},n_{23},n_{33}).
$$
To provide the classification, it is sufficient to consider the options:
$\quad 1)\quad a_1 \ne 0, \quad 2)\quad a_3 \ne 0, \quad a_1 = a_2 = 0. $ \quad

\quad

a) $ a_1\ne 0 \Rightarrow \beta_a=\frac{\alpha_a \beta_1}{\alpha_1}$. 
 $$
 \hat{W}_I \hat{n}_\alpha = (\hat{\omega}_\beta -\hat{Q}_1 \hat{n}_\beta) \Rightarrow
 \hat{n}_\alpha = \hat{W}_I^{-1}(\hat{\omega}_\beta -\hat{Q}_1 \hat{n}_\beta) ,$$
 $$
\beta_1 \hat{W}_I \hat{n}_\alpha = \alpha_1\hat{\omega}_\alpha -\beta_1\hat{Q}_1 \hat{n}_\beta  \Rightarrow \beta_1\hat{\omega}_\beta - \alpha_1\hat{\omega}_\alpha  =0  \Rightarrow
 $$
 \begin{equation}\label{18c}
   \left\{\begin{array}{rl}
\alpha_1\dot{\alpha}_2 +\beta_1\dot{\beta}_2 =0, \\
\alpha_1\dot{\alpha}_3 +\beta_1\dot{\beta}_3 =0, \\
\alpha_1\dot{\alpha}_1 +\beta_1\dot{\beta}_1 = 0. \\
\end{array}\right.  \Rightarrow \left\{\begin{array}{rl}
\alpha_1 = e \sin\varphi, \quad\beta_1 = e \cos\varphi, \quad e = const, \\
\alpha_2 = e c_2 \sin\varphi, \quad \beta_1 = e c_2 \cos\varphi, \quad e, c_2 = const, \\
\alpha_3 = e c_3 \sin\varphi, \quad \beta_1 = e c_3 \cos\varphi, \quad e, c_3 = const. \\
\end{array}\right.\end{equation}
Here:
$$
 \hat{W}_I = \begin{pmatrix} \alpha_1 & \alpha_2 & \alpha_3 &  \\ 0 &  \alpha_1 & 0
\\0 & 0 & \alpha_1
 \end{pmatrix}, \hat{W}_I^{-1}= \begin{pmatrix} \frac{1}{\alpha_1} & -\frac{\alpha_2}{\alpha_1^2} & -\frac{\alpha_3}{\alpha_1^2} &  \\ 0 &  \frac{1}{\alpha_1} & 0
\\0 & 0 & \frac{1}{\alpha_1}
 \end{pmatrix}, \hat{Q}_I = \begin{pmatrix} 0 & 0 & 0 &  \\ \alpha_2 &  \alpha_3 & 0
\\0 & \alpha_2 & \alpha_3,
 \end{pmatrix}
 $$
$$
\alpha_1 = e \sin\varphi, \quad\beta_1 = e \cos\varphi, \quad e, c_a = const,
$$
Then matrices $\hat{W}_I, \hat{W}_I^{-1}, \hat{Q}_I$ and lines $\hat{\omega}^T$ take the form:
$$
 \hat{W}_I = \sin{\varphi}\hat{P},\quad \hat{W}_I^{-1}= \frac{1}{\sin{\varphi}}\hat{P}^{-1} ,\quad \hat{Q}_I = \sin{\varphi}\hat{Q}.
 $$
 $$
 \hat{P}=\begin{pmatrix} 1 & c_2 & c_3 &  \\ 0 &  1 & 0
\\0 & 0 & 1
 \end{pmatrix}, \quad
 \hat{P}^{-1} = \begin{pmatrix} 1 & -c_2 & -c_3 \\ 0 &  1 & 0
\\0 & 0 & 1 \end{pmatrix},\quad
\hat{Q} = \begin{pmatrix} 0 & 0 & 0 &  \\ c_2 &  c_3 & 0
\\0 & c_2 & c_3,
 \end{pmatrix}
 $$
$$
 \hat{\omega}^T_\alpha = \hat{\omega}^T_\beta = \dot{\varphi}\sin{\varphi} \hat{C}^T = \dot{\varphi}\sin\varphi( 1, c_2,c_3),
 $$
$$
\hat{n}_\alpha = \hat{w}^{-1}(\dot{\varphi}\hat{C}^T -\hat{Q} \hat{n}_\beta)
$$
Function \quad $n_{22}, n_{23}, n_{33},$ \quad as well as the function \quad $\varphi$ \quad are arbitrary functions of \quad $u^0$.

\quad

 b) $V_a=0, \quad \alpha_3 \ne 0. \quad  \Rightarrow \beta_1=\beta_2 =0$.\quad The system of Maxwell equations has the form:
$$
\alpha_3 n_{13}=\alpha_3 n_{23}=0,\quad \alpha_3 n_{33}=-\dot{\beta}_3, \quad \beta_3 n_{33}=\dot{\alpha}_3.
$$
$$
a_3 \dot{a}_3 + \beta_3\dot{\beta}_3=0 \Rightarrow a_3 =c\sin{\varphi}, \quad \beta_3 = \cos{\varphi}
$$
From here:
$$
 n_{33}=\dot{\varphi}, \quad n_{13}= n_{23}=\alpha_1 = \alpha_2  =\beta_1 = \beta_2 = 0,\quad \alpha_3 = c \sin\varphi, \quad \beta_3 = c\cos\varphi. $$
Functions \quad $ \varphi,\quad n_{11}, \quad  n_{12}, \quad  n_{22}$ - are arbitrary functions on \quad $u^0, \quad c=const$.

\section{Conclusion}
It is known that homogeneous spaces of $IV$ and $IX$ types according to Bianchi classification include as special cases the spaces of constant curvature.This causes a special interest to them in cosmology. In the Universe with the metric of homogeneous space all physical fields are invariant with respect to the group of motions of the space-time. Therefore, exactly such fields should be considered in the first place when solving the self-consistent Einstein equations, in particular the Einstein-Maxwell equations. The final goal of classification of PSS with admissible electromagnetic fields is to enumerate all electrovacuum solutions of the Einstein-Maxwell equations. In \cite{34}, \cite{35} the complete classification of vacuum solutions of the Maxwell equations for homogeneous spaces with solvable groups of motions has been carried out. In the present paper the same problem is solved for HPSS of $IX$-type. For the final decision of the first stage of the classification problem it remains to consider HPSS $VIII$-type, which will be done in the next paper. The results obtained will be used in the second stage for integration of the corresponding Einstein-Maxwell equations.

\quad

FUNDING: The work is supported by Russian Science Foundation, project number N 23-21-00275.

INSTITUTIONAL REVIEW BOARD STATEMENT: Not applicable.

INFORMED CONSENT STATEMENT: Not applicable.

DATA AVAILABILITY STATEMENT: The data that support the findings of this study are available within the article.

CONFLICTS OF INTEREST: The author declares no conflict of interest.

\end{document}